\documentclass[letter]{aa}                            

\usepackage{graphicx}
\usepackage{array}
\usepackage{lscape}                                
\usepackage{txfonts}
\usepackage{natbib}
\usepackage{longtable}
\usepackage{color}

\newcommand{\kms} {km\,s$^{-1}$}
\newcommand{\vsini} {$v$\,sin\,$i$}
\newcommand{\Teff} {T$_{\rm eff}$}
\newcommand{\grav} {log\,{\em g}}

\begin{document}


%

\title{The IACOB project\thanks{Based on observations made with (1) the Nordic Optical 
        Telescope, operated
       on the island of La Palma jointly by Denmark, Finland, Iceland,
       Norway, and Sweden, in the Spanish Observatorio del Roque de los
       Muchachos of the Instituto de Astrofisica de Canarias, and (2) the 
       European Southern Observatory Very Large Telescope in programme 182.D-0222.}
        }

\subtitle{II. On the scatter of
O-dwarf spectral type -- effective temperature calibrations} 

\author{S. Sim\'on-D\'{\i}az\inst{1,2}, A. Herrero\inst{1,2}, C. Sab\'in-Sanjuli\'an\inst{3}, F. Najarro\inst{4},
M. Garcia\inst{4}, J. Puls\inst{5}, N. Castro\inst{6}, C. J. Evans\inst{7}}

\institute{Instituto de Astrof\'isica de Canarias, E-38200 La Laguna, Tenerife, Spain              
             \and
             Departamento de Astrof\'isica, Universidad de La Laguna, E-38205 La Laguna, Tenerife, Spain
             \and
             Departamento de F\'isica, Universidad de La Serena, Av. Cisternas 1200 Norte, La Serena, Chile
             \and
             Centro de Astrobiología, CSIC-INTA. Ctra. Torrejón a Ajalvir km.4, E-28850 Torrejón de Ardoz, Madrid, Spain
             \and
             Universitäts-Sternwarte, Scheinerstrasse 1, 81679, München, Germany
             \and 
             Argelander Institut für Astronomie, Auf den Hügel 71, 53121, Bonn, Germany
             \and
             UK Astronomy Technology Centre, Royal Observatory, Blackford Hill, Edinburgh, EH9 3HJ, UK}
           
\offprints{ssimon@iac.es}

\date{Submitted/Accepted}

\titlerunning{On the scatter of O-dwarf SpT---\Teff\ calibrations}

\authorrunning{S. S.-D. et al.}

%
\abstract{We are now in an era of large spectroscopic surveys of
  OB-type stars. Quantitative spectroscopic analysis of these modern
  datasets is enabling us to review the physical properties of blue
  massive stars with robust samples, not only revisiting mean
  properties and general trends, but also incorporating information
  about the effects of second-order parameters.}  {We investigate the
  spectral type -- effective temperature (SpT\,--\,\Teff) calibration
  for O-type dwarfs and its claimed dependence on metallicity, using
  statistically meaningful samples of stars extracted from the IACOB
  and VFTS surveys.}  {We performed a homogeneous differential
  spectroscopic analysis of 33 Galactic and 53 LMC O~dwarfs (spanning
  spectral types of O4\,--\,O9.7) using the {{\sc iacob-gbat}} package,
  a $\chi^2$-fitting algorithm based on a large pre-computed grid of
  {\sc fastwind} models, and standard techniques for the
  hydrogen/helium analysis of O-type stars.  We compared the estimated
  effective temperatures and gravities as a function of (internally
  consistent) spectral classifications.}  {While the general trend is
  that the temperature of a star increases with earlier spectral types
  and decreasing metallicity, we show that the wide range of
  gravities found for O-type dwarfs -- spaning up to
  0.45\,-\,0.50\,dex in some spectral bins -- plays a critical role on
  the dependence of the effective temperature calibrations as a
  function of spectral type and metallicity.}  {This result warns us
  about the use of SpT\,-\,\Teff\ calibrations for O dwarfs that
  ignore the effects of gravity, and highlights the risks of employing
  calibrations based on small samples. The effects of this scatter in
  gravities (evolutionary status) for O-type dwarfs should be included
  in future recipes that employ SpT\,-\,\Teff\ calibrations.}
%
\maketitle
%





%

\section{Introduction}\label{section1}

An important deliverable that studies of massive stars must provide to
the broader astrophysics community is a series of recipes in which the
physical properties of OB-type stars (such as effective temperature,
gravity, luminosity, radius, mass, wind momentum, and number of ionizing
photons) are calibrated against more easily determined quantities
(e.g., spectral type and luminosity class), or against the input
parameters used in population synthesis models (mass and age). These
recipes are expected to be applied to the study of different
environments across the Universe, therefore characterizing the
effect of metallicity ($Z$) on these calibrations is of key
importance.

The spectral type -- effective temperature (SpT\,--\,\Teff)
calibration for O-type stars has been extensively studied, across a
range of metallicities, and using the most advanced codes
available \citep[see, for example,][]{Vac96, Mar02, Her02, Bou03, Rep04,
  Mas04, Mas05, Mar05, Hea06, Mok07, Mas09, Riv12, Gar13}. These and 
previous studies have generally investigated the SpT\,--\,\Teff\ calibrations
separated into three luminosity classes (LC): dwarfs (V), giants
(III), and supergiants (I). The global picture is clear in terms of
the dependence of \Teff\ on SpT and LC: early O-type stars are
hotter than late-O stars and dwarfs are hotter than supergiants. The dependence
on $Z$ is less clear; while there are some hints that (as expected)
stars are hotter at lower $Z$ (for a given SpT), this is not always supported
by the observations \citep{Mas04, Mar05}. In addition, there is another
important factor that has not been sufficiently explored to date -- the
remarkable scatter found in \Teff\ for a given SpT\,$+$\,LC\,$+$\,$Z$ combination.

Past studies have (necessarily) drawn their conclusions from
relatively small samples of stars and, in some cases, from the
comparison of results obtained using different codes and techniques. In
this context, the efforts of the GOSSS, IACOB, and OWN projects
\citep[][respectively]{Mai11, Sim11a, Bar10} in the Milky Way, and the
VLT-FLAMES Tarantula Survey \citep[VFTS,][]{Eva11} in the 30~Doradus
region of the Large Magellanic Cloud provide us with an excellent
opportunity to improve this situation.  For the first time,
large samples of O stars in two different metallicity environments are
being investigated homogeneously using the same codes and techniques.
This not only concerns the quantitative analysis, but also the
spectral classification \citep[][]{Sot11, Wal14}. 

Quantitative spectroscopic analysis of the IACOB, OWN, and VFTS O-type
samples is on-going and will be published elsewhere \cite[see, e.g., ][]{Sab14, Bes14}. In this letter we present first results on
the investigation of the SpT\,--\Teff\ calibration in O dwarfs and its
claimed dependence on $Z$ using two carefully selected sub~samples of
stars from the IACOB database and the VFTS.

\section{Observations and analysis}\label{section2}

The Galactic sample is drawn from the class V stars observed by the
IACOB project \cite[see][]{Sim14}, excluding objects that were detected as
double-lined or large-amplitude binaries. The final sample
comprises 33 O-type dwarfs, spanning a range in SpT from O4 to O9.7.
To investigate the effects of $Z$ we considered O-type dwarfs from the
VFTS \cite[again omitting any target with indications of binarity or a
composite spectrum, see][]{San13, Wal14, Sab14}.  The LMC sample
comprises 53 O-type dwarfs\footnote{To ensure that the range of SpTs
is the same between the two samples, the O2-/O3-type stars
observed by the VFTS were not considered.}, which are essentially
those listed in Tables~A.1 and A.2 from \citet{Sab14}, but without
stars with uncertain SpTs, or indications
of more than one target in the fibre.

Spectral types of the Galactic and LMC samples were taken from
\cite{Sot11} and \cite{Wal14}, respectively.  The stellar and wind
parameters of these two samples were determined using the {\sc iacob-gbat}
package \citep{Sim11b}, based on a $\chi^2$-fitting algorithm applied
to a large pre-computed grid of {\sc fastwind} \citep{San97, Pul05} models,
and standard techniques for the hydrogen/helium analysis of O-type stars
\citep[see, e.g.,][]{Her92, Rep04}. The complete spectroscopic analysis
of the IACOB sample will be presented in a forthcoming paper
(Sim\'on-D\'iaz et al., in prep.). Details of the {\sc iacob-gbat}
analysis of the VFTS O-dwarf sample were given by \cite{Sab14}.
Estimates for the projected rotational velocities (\vsini) used in the
analyses are those given by \citet[][IACOB]{Sim14} and
\citet[][VFTS]{Ram13}.

Before we discuss our results, we stress that we removed
known binaries and composite spectra, we performed a homogeneous 
differential analysis of both samples, and the spectral classification
of all the stars considered was (self-consistently) undertaken
by the same team.

\section{Results and discussion}\label{section3}

%

\begin{figure}[!ht]
\centering
\includegraphics[width=6.5cm,angle=90]{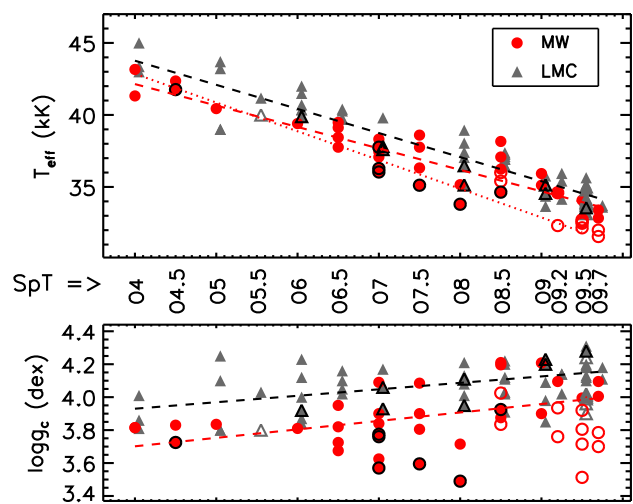}
\caption{\Teff\ and \grav$_{\rm c}$ estimates for the Galactic (red circles) and LMC (gray triangles)
  O-type dwarfs as a function of SpT; corresponding
  linear fits are overplotted with red and black dashed
  lines.  Stars with \vsini\,$>$\,250\,\kms\ are highlighted in black.  
  Luminosity class IV stars are plotted as red and gray open
  symbols and were not used for the linear fits (see text). The widely
  used SpT\,--\,\Teff\ calibration by \cite{Mar05} is also shown for reference
  (red dotted line).}
\label{f1}
\end{figure}
%


\begin{table*}[t!]
\begin{center}
  \caption{\footnotesize Summary of \Teff\ and \grav\ results for the IACOB and VFTS O-dwarf (V) samples.}
\label{t1}
\begin{tabular}{r c c c c c c c c r c c c c c c c} 
\hline
\noalign{\smallskip}
\cline{1-8} \cline{10-17}  
\noalign{\smallskip}
SpT & \# stars &   & \multicolumn{2}{c}{\Teff\ [kK]} & & \multicolumn{2}{c}{\grav$_c$\ [dex]} & & SpT & \# stars & & \multicolumn{2}{c}{\Teff\ [kK]} & & \multicolumn{2}{c}{\grav$_c$\ [dex]}  \\ 
\cline{1-2} \cline{4-5} \cline{7-8} \cline{10-11} \cline{13-14} \cline{16-17}
\noalign{\smallskip}
\multicolumn{2}{c}{MW (IACOB)} & & Mean & Range$^{1}$ & & Mean & Range & & \multicolumn{2}{c}{30~Dor (VFTS)} & & Mean & Range & & Mean & Range  \\ 
\cline{1-8} \cline{10-17}  
\noalign{\smallskip}
O4   &  2 & & 42.2 &   1.8 &  &  3.82 &  0.00 &  & O4   & 3 & & 43.8 &   2.0 &  &  3.89 &  0.20 \\ 
O4.5 &  2 & & 42.1 &   0.6 &  &  3.78 &  0.11 &  & O4.5 & 0 & &  $-$ &   $-$ &  &   $-$ &  $-$  \\ 
O5   &  1 & & 40.4 &   $-$ &  &  3.84 &  $-$  &  & O5   & 4 & & 41.2 &   4.7 &  &  3.99 &  0.45 \\ 
O5.5 &  0 & &  $-$ &   $-$ &  &  $-$  &  $-$  &  & O5.5 & 1 & & 41.2 &   $-$ &  &  4.03 &  $-$  \\ 
O6   &  1 & & 39.4 &   $-$ &  &  3.81 &  $-$  &  & O6   & 6 & & 40.8 &   2.1 &  &  4.00 &  0.39 \\ 
O6.5 &  4 & & 38.7 &   1.7 &  &  3.79 &  0.28 &  & O6.5 & 4 & & 40.0 &   0.7 &  &  4.08 &  0.14 \\ 
O7   &  7 & & 37.0 &   2.3 &  &  3.79 &  0.52 &  & O7   & 3 & & 38.4 &   2.2 &  &  4.05 &  0.24 \\ 
O7.5 &  4 & & 36.9 &   3.5 &  &  3.85 &  0.49 &  & O7.5 & 0 & &  $-$ &   $-$ &  &   $-$ &  $-$  \\ 
O8   &  2 & & 34.5 &   1.4 &  &  3.60 &  0.22 &  & O8   & 6 & & 37.2 &   3.8 &  &  4.08 &  0.26 \\ 
O8.5 &  4 & & 36.5 &   3.5 &  &  4.05 &  0.34 &  & O8.5 & 7 & & 36.9 &   0.9 &  &  4.09 &  0.31 \\ 
O9   &  2 & & 35.5 &   0.8 &  &  4.05 &  0.31 &  & O9   & 5 & & 34.7 &   2.1 &  &  4.07 &  0.38 \\ 
O9.2 &  1 & & 34.6 &   $-$ &  &  4.09 &  $-$  &  & O9.2 & 3 & & 35.2 &   1.8 &  &  4.10 &  0.12 \\ 
O9.5 &  1 & & 34.1 &   $-$ &  &  3.99 &  $-$  &  & O9.5 & 9 & & 34.5 &   2.4 &  &  4.19 &  0.21 \\ 
O9.7 &  2 & & 33.1 &   0.6 &  &  4.05 &  0.09 &  & O9.7 & 2 & & 33.7 &   0.1 &  &  4.14 &  0.07 \\ 
\cline{1-8} \cline{10-17}  
\noalign{\smallskip}
\hline
\end{tabular}
\tablefoot{Typical uncertainties in \Teff\ for the VFTS sample are in
  the range $\pm$0.5\,--\,1.5~kK, while the
  formal errors in \grav\ range between 0.07 and 0.15\,dex
  \citep[however, systematic errors make 0.10\,dex a more reasonable
  lower limit in $\Delta$\grav, see][]{Sab14}. Typical uncertainties
  in results for the IACOB sample are on the same order or slightly
  smaller in some cases.$^{(1)}$ Difference between
  the highest and lowest values.}
\end{center}
\end{table*}

The \Teff\ and \grav$_{\rm c}$ (gravity~corrected for centrifugal
acceleration) estimates for the Galactic and LMC samples are shown as a
function of SpT in Figure~\ref{f1}; the SpT\,--\,\Teff\ and
SpT\,--\,\grav\ calibrations from linear fits to the data are
overplotted as dashed lines. These results are also summarized in
Table~\ref{t1}, in which we indicate the number of stars and the means
and ranges of the \Teff\ and \grav$_{\rm c}$ estimates per SpT bin.

The SpT\,--\,\Teff\ calibrations agree well with previous
studies, in which LMC O-type dwarfs ($Z$\,$\sim$\,0.5\,$Z_{\odot}$) are
$\sim$\,1000\,--\,2000\,K hotter than Galactic stars
\citep[e.g.,][]{Mok07,Riv12}.  However, one important feature of our
results 
\citep[which is also present in other previous works dealing with 
SpT\,--\,\Teff\ calibrations, see, e.g.,][]{Vac96, Mar05, Mas09}
is the non-negligible 
scatter of \Teff\ for most of the SpT
bins. This dispersion -- of up to $\sim$3500\,K in the more extreme
cases -- is higher than the estimated uncertainties from the
quantitative spectroscopic analysis (typically $\pm$\,500\,--\,1500\,K,
depending on the specific example).  As a consequence, we can find (1)
stars with same metallicity, different SpTs, but the same \Teff, (2) stars
with different metallicities, the same SpT, but the same \Teff, and (3)
Galactic O stars with a higher \Teff\ than those in the LMC for a
given SpT. Thus, the situation is more complex than simply comparing
two linear SpT\,--\,\Teff\ calibrations resulting from samples
with different metallicities.

This situation is easily understood when one also considers the
gravities in the interpretation of the results. As illustrated by
Fig.~\ref{f1} and Table~\ref{t1}, there is also a broad scatter
in the gravities associated with a given SpT bin (up to 0.45\,--\,0.50
dex in the most extreme cases). In particular, we note that the
Galactic O-type dwarfs span a range of gravities of 4.2 to
3.5~dex\footnote{The slightly wider range in \grav$_{\rm c}$ found in
the IACOB sample is due to a few mid O-type stars with \vsini\ above
250~\kms. In these cases both the spectral classification and the
spectroscopic analysis is less accurate than for the rest of the
sample.}, with a range of 4.3 to 3.8~dex for the LMC stars.

From our results it is clear that adopting a unique gravity for the
O dwarfs is an oversimplified assumption. As a consequence, the
dispersion found in the SpT\,--\,\Teff\ calibrations is to be
expected and arises from the degeneracy between \Teff\ and \grav\ to
reproduce a similar He ionization equilibrium (i.e., SpT) for a given
star. In this context, we note that there is a {\em rough}
correlation between the scatter found in \Teff\ and \grav, and between
the low/high \Teff\ and low/high \grav\ cases in each bin. 

This effect is known and is the reason that the \Teff\ of a
supergiant is lower than that of a dwarf with the same SpT. Indeed, it
is also the reason why separate SpT\,--\,\Teff\ calibrations are
normally given for different luminosity classes
\citep[e.g.,][]{Mar05}.  
However, given the wide range of gravities found for stars classified as O dwarfs,
the most important insight from Fig.~\ref{f1} is that assuming
a unique (but metallicity-dependent) SpT\,--\,\Teff\ calibration for stars of this 
luminosity class is an oversimplified recipe.
Indeed, the situation is even more complex when one also considers
the effects that rotation and resolving power produce on the
classification of O-type spectra \cite[see][]{Mar11}.

\begin{figure}[!t]
\centering
\includegraphics[width=8.cm,angle=0]{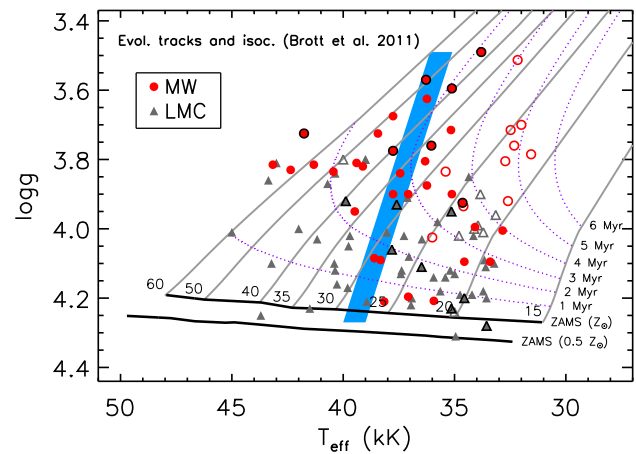}
\caption{Galactic and LMC O-dwarf results plotted in the
  \Teff-\grav\ diagram (same symbols as Fig.~\ref{f1}).  Evolutionary
  tracks (solid grey lines, with initial masses as indicated at the
  ZAMS) and isochrones (dotted purple lines) are from \citet{Bro11}
  for solar metallicity; the ZAMS for both Z\,=\,Z$_{\odot}$ and
  0.5~Z$_{\odot}$ are indicated as black lines. The blue region
  indicates where the O7~V stars are located.}
\label{f2}
\end{figure}

The location of the two samples in the \Teff-\grav\ diagram\footnote{From here 
on, \grav\ refers to the actual gravity, denoted previously by \grav$_{\rm c}$ }
is shown in Fig.~\ref{f2} together with evolutionary tracks and isochrones
from \cite{Bro11} for $Z$\,$=$\,$Z_{\odot}$ and $v_{\rm rot,ini}$\,$=$\,220\,\kms\ 
and the position of the zero-age main
sequence (ZAMS) for $Z$\,$=$\,0.5\,$Z_{\odot}$ (same physics and
$v_{\rm rot, ini}$). For reference, the blue band indicates the
approximate region where O7~V stars are located. As expected from the
\Teff\,--\grav\ degeneracy in reproducing a similar He ionization
equilibrium, this band is tilted and not vertical, with a slope
of about 750\,K per 0.1\,dex in \grav.  The figure also shows the
displacement of the ZAMS toward higher \Teff\ and \grav\ with decreasing
metallicity \citep[see][for more details]{Mow98}. This effect,
combined with the effects of line blanketing \citep{Mar02, Mar05,
Rep04}, leads to the SpT\,--\,\Teff\ and SpT\,--\grav\
calibrations in the LMC to be different to those in the Galaxy. Fig.~\ref{f1}
shows that this is the observed trend when the analyzed samples are
statistically meaningful; however, as we indicate in
Sect.~\ref{section4}, one must handle this argument with care when
dealing with an individual star or when extracting conclusions
from the study of small samples.

%
%

In Figs.~\ref{f1} and \ref{f2} we also included results for a
small number of class IV (sub~giant) stars.  While the general trend
is to find the O~IV stars in the lower \Teff\ and lower \grav\
envelope of the distribution, it is noteworthy that some of them are
located in the same region as the O-type dwarfs.  In many cases,
particularly when the quality of the observations is poor, it
can be difficult to morphologically separate O~IV from O~V stars .
Moreover, it may also be difficult to distinguish the
two classes from the results of the spectroscopic analysis because the
estimated gravities for the O~IV stars (which are normally
concentrated amongst the late-O spectral types) are compatible with the
lower end of the \grav\ distribution found for mid- and early-O dwarfs.
Hence, by misclassifying O~IV stars as O dwarfs, or if one increases
the statistics by combining stars from both luminosity classes, the
scatter in the SpT\,--\,\Teff\ calibration due to gravity effects
would become even larger.

%
\begin{figure}[!t]
\centering
\includegraphics[width=8.0cm,angle=0]{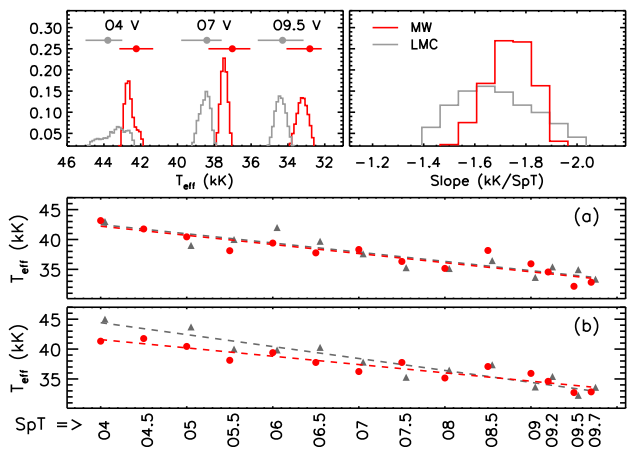}
\caption{{\em Upper panels:} Results from a Monte-Carlo simulation
  where one star is drawn from the Galactic and LMC samples per
  SpT bin. Probability distributions for the \Teff\ of three
  representative SpT bins and for the overall slope of the calibration
  are shown in the left- and right-hand panels. {\em
    Lower two panels:} Two possible SpT\,--\,\Teff\ relations arising
  from the simulations (employing the same symbols as in
  Fig.~\ref{f1}), highlighting the potential consequences of drawing
  conclusions from small samples.}
\label{f3}
\end{figure}
%

\section{Implications}\label{section4}

In the following, we provide some important consequences (mainly warnings) 
implied by our results:

\begin{enumerate}
\item Given the significant scatter of \Teff\ associated with each
  SpT bin, one must be careful when drawing conclusions from the
  analysis of small samples (either for a given $Z$ or when comparing
  results across a range of environments). To illustrate this, we present a comparison of results 
  from a Monte-Carlo simulation in Fig.~\ref{f3} in which a smaller sample (just one star per SpT bin) is
  drawn from each global sample (including the O~IV stars).  The upper
  panels show the probability distributions for the \Teff\ associated
  with three SpT bins (left panel, in which the means and ranges,
  including the O~IV stars, are also indicated), and for the derived
  slope of the calibrations (from linear fits to the data).  The lower
  panels show two possible SpT\,--\,\Teff\ relations drawn from such
  simulations in which (a) the same calibrations would be obtained
  for the two metallicities; (b) a different slope is obtained, in
  which $\Delta$\Teff\,$\sim$\,4\,000\,K is found for the hottest
  stars, but where similar temperatures are associated with the
  late-O stars.

  \item We compare in Fig.~\ref{f1} our linear SpT\,--\,\Teff\ calibration for Galactic
O dwarfs with the one suggested by \citet[][MSH05]{Mar05}, which is widely used. 
Obviously, the agreement is far from being
satisfactory in the late-O regime, but the discrepancy can be
explained by taking into account (i) gravity and small-number
statistics effects and (ii) the fact that MSH05 included a few O3\,-\,O3.5~V 
stars in their linear fit, while \cite{Riv12}
have indicated a change in slope at SpT around O4. Thus, the
calibration by MSH05 might suggest too low \Teff\ values in the late
O-dwarf regime.

\item MSH05 also proposed a characteristic constant gravity of 3.92 dex
 for all O dwarfs. We showed that this is a dangerous oversimplification.
 In fact, as we discuss in item 5, any possible attempt to provide a
 SpT\,--\,\grav\ calibration will critically depend on the age and mass 
 distribution (plus other effects such as the \vsini\ of the stars) of the analyzed O-star sample.

\item The broad range in gravities not only affects the \Teff\
  associated with an O dwarf for a given SpT, but also the
  corresponding stellar mass and radius (and consequently other related
  quantities such as the number of ionizing photons). For example,
  Fig.~\ref{f2} shows that a Galactic O7~V star may have a range in
  (evolutionary) mass of about 10~M$_{\odot}$. From simple
  calculations, the radius of an O7~V star with \grav\,$=$\,3.7 will be
  roughly twice that of a star on the ZAMS (\grav\,$\sim$\,4.2).
  In addition, an O7~V star close to the ZAMS will only emit
  roughly half as many ionizing photons as the star with a lower
  gravity.  This is counterintuitive because the ZAMS star is hotter (by $\sim$3,500\,K), but the combined
  effect of \Teff\ and radius produces this result. Of course, these
  numbers are purely for illustration and are based on evolutionary
  calculations (and estimated for a unique initial \vsini).  A more
  thorough investigation of the scatter in stellar mass, radius, and
  number of ionizing photons, using actual luminosities derived from
  absolute magnitudes, is one direction for future work.

\item We highlighted the importance of gravity on the
  interpretation of the scatter seen in SpT\,--\,\Teff\ calibrations
  for O dwarfs, but our comments can be 
  directly translated into any parameter affecting the evolutionary 
  status (governed by mass, age, initial rotational velocity, mass-loss, etc.~) 
  of the individual stars in the considered sample. For example, in a 
  sample drawn from a single-age starburst population, all stars with masses above 15
  M$_{\odot}$ will only have the same gravity if it is younger than
  1\,Myr; as soon as the population evolves, the early-O dwarfs will
  have lower gravities than the late-O stars 
  (as illustrated by the isochrones presented in Fig.~\ref{f2}).
  As a consequence, the SpT\,--\,\Teff\ calibration that should be employed in these two
  situations is different (both in terms of absolute values and
  slope): in general terms, an O dwarf with a given SpT
  will be hottest when it is on the ZAMS and will cool as it evolves
  during the following 3-4 Myr, during which time it would still be
  classified as an O dwarf.

\end{enumerate}
%
\begin{acknowledgements}
This work has been funded by the Spanish Ministry of Economy and
  Competitiveness (MINECO) under the grants AYA2010-21697-C05-01/04, AYA2012-39364-C02-01, 
  FIS2012-39162-C06-01, ESP2013-47809-C3-1-R, and Severo Ochoa SEV-2011-0187, and 
  by the Canary Islands Government under grant PID2010119.
\end{acknowledgements}



%
\end{document}